\documentclass[sn-mathphys-num]{sn-jnl}

\usepackage{bm,amssymb,amsmath,dsfont,mathrsfs,amstext,latexsym,physics,relsize}
\usepackage{graphicx}

\usepackage[usenames,dvipsnames]{xcolor}
\hypersetup{colorlinks=true,citecolor=blue,urlcolor=blue,linkcolor=blue}

\newcommand{\figpanel}[2]{\hyperref[#1]{\ref{#1}#2}}

\begin{document}

\title[Universal Defect Statistics in Counterdiabatic Quantum Critical Dynamics]{Universal Defect Statistics in Counterdiabatic Quantum Critical Dynamics}

\author*[1]{\fnm{Andr\'as} \sur{Grabarits}}
\email{andras.grabarits@uni.lu}

\author[1,2]{\fnm{Adolfo} \sur{del Campo}}

\affil[1]{\orgdiv{Department of Physics and Materials Science}, \orgname{University of Luxembourg}, \orgaddress{\city{Luxembourg}, \postcode{L-1511}, \country{Luxembourg}}}

\affil[2]{\orgname{Donostia International Physics Center}, \orgaddress{\city{San Sebasti\'an}, \postcode{E-20018}, \country{Spain}}}

\abstract{
Counterdiabatic driving  (CD) provides a framework for suppressing excitations in nonadiabatic processes. Exact CD protocols require nonlocal control fields, and  CD approximations with tailored locality are needed for their implementation. However, the performance of local CD schemes remains poorly understood.
Here, we develop an analytically tractable local CD expansion scheme and establish a universal scaling theory governing the defect statistics after crossing a quantum phase transition as a function of the CD locality order. Our predictions are tested on the transverse field Ising model and long-range Kitaev models. Our results provide an analytical framework for evaluating the effectiveness of local CD protocols in quantum state preparation, control, and optimization.
}

\maketitle

\section*{Introduction}

Adiabatic protocols play a key role in quantum control, but their implementation generally requires slow driving. In certain scenarios, they are thus not applicable. This is the case when driving a continuous quantum phase transition (QPT), characterized by the closing of the gap between the ground and the first excited
state at the critical point. Across a QPT, adiabaticity breaks down, giving rise to defects as excitations \cite{Sachdev2011Quantum,Kato1950Adiabatic,Jansen2007Bounds}. Yet, adiabatically crossing QPTs at a finite driving rate is desirable for a range of important applications, such as quantum state preparation in quantum simulation and adiabatic quantum computation~\cite{Cirac2012,Albash_AQC_2018}.
A variety of approaches have been investigated for defect suppression. They involve quantum control \cite{Doria_Optcontr2011,WuNanduri15,Mao24}, nonlinear driving \cite{Diptiman08,Barankov08}, inhomogeneous or local driving \cite{Dziarmaga_2010,Collura_2010,GomezRuiz19}, multi-parameter driving \cite{SauSengupta14}, a symmetry-breaking bias \cite{Rams19,Kormos20}, and coupling to a tailored environment \cite{Dora2019,King2023,Bacsi2023,Mahunta25}. Defect suppression is also enhanced in specific scenarios that involve the crossing of gapless lines and multicritical points \cite{Divakaran08,DengOrtizViola09} and broken chiral symmetry \cite{Yan21}.

A particularly powerful and versatile approach relies on assisting the time evolution with additional control fields through counterdiabatic driving (CD)~\cite{Demirplak2003Adiabatic, Demirplak2005Assisted,Demirplak2008Consistency,Berry2009Transitionless}. The auxiliary CD fields are highly nonlocal in many-body systems, making their experimental implementation in quantum devices challenging \cite{delcampo12,Takahashi2013Transitionless,Damski2014Counterdiabatic}. This motivates the search for approximate CD schemes with controlled locality to ease their implementation. 
Local CD protocols can be designed using variational methods \cite{Saberi2014Adiabatic,PolkovnikovSels_2017} and truncated expansions in Krylov space~\cite{Claeys2019Floquet,Takahashi2024Shortcuts,Bhattacharjee2023Lanczos}.
Combined with digital quantum simulation, these approaches open a realistic route toward implementation, and defect suppression across a QPT has recently been demonstrated by Trotterizing the dynamics in the circuit model for quantum computation \cite{Visuri2025}. Exact CD protocols with nonlocal control fields reproduce adiabatic dynamics and can completely suppress defect formation. By contrast, the performance of approximate local CD protocols remains poorly understood.

In this work, we present exact analytical predictions and numerical evidence for the universal defect statistics generated during the CD-assisted crossing of a QPT. We develop local CD expansion schemes that remain analytically tractable for arbitrary solvable critical models. Using this framework, we show that the defect cumulants decay according to universal power laws with the locality order, and the full defect statistics approaches a Gaussian distribution. These features are demonstrated through exact results in the transverse-field Ising model (TFIM), where our expansion coincides with the Krylov subspace construction. We further corroborate the universal scaling laws in the long-range Kitaev models (LRKM)---both within the Krylov subspace and our novel expansion framework. These results provide a general framework for assessing the efficiency of local CD protocols in quantum control, state preparation, and optimization.

\section*{Results}
\subsection*{Defect statistics with local counterdiabatic driving}

Consider the adiabatic trajectory generated by an uncontrolled system Hamiltonian $H[g(t)]$. CD makes it possible to fast-forward such evolution at an arbitrary rate by assisting the dynamics with the (exact) CD term so that the total Hamiltonian reads~\cite{Demirplak2003Adiabatic,Demirplak2005Assisted,Demirplak2008Consistency, Berry2009Transitionless},
    $H_\mathrm{CD}(t) = H[g(t)] + H_1[g(t)]$, where 
    $H_1[g]=i\,\dot g  \sum_n(|\partial_gn\rangle\langle n|-|n\rangle\langle n|\partial_g n\rangle\langle n|)$.
 This Hamiltonian becomes highly nonlocal in many-body systems, in particular across a QPT, making its experimental implementation challenging \cite{delcampo12,Saberi2014Adiabatic}.
To address this issue, a feasible strategy is provided by the nested-commutator expansion with a controlled locality \cite{Claeys2019Floquet}, which has traditionally been captured within the Krylov subspace expansion and by the variational principle.

However, these approaches typically remain analytically tractable only for simple single-particle settings or for the first few expansion orders, even in otherwise simple many-body models. This makes it difficult to extract general or universal features of counterdiabatic quantum critical dynamics.
To overcome this limitation, we first establish a new local CD expansion scheme within an orthogonal operator subspace, retaining analytical control for arbitrary solvable critical systems.
In particular, consider the time-dependent Hamiltonian
\begin{eqnarray}\label{eq: H_0}
H(t)=\sum_{r,l}\left(j_rc_lc^\dagger_{l+r}+d_rc_lc_{l+r}+\mathrm{h.c.}\right)-g(t)\sum_lc^\dagger_lc_l,
\end{eqnarray}
with $c_l,\,c^\dagger_l$ denoting fermionic creation and annihilation operators. The Fourier transforms $j_k=\sum_r \cos(kr)j_r\propto\cos\varphi_k$ and $d_k=\sum_r\sin(kr)d_r\propto\sin\theta_k$, with $\theta_{k\ll1}\sim k^{\beta-1},\,\varphi_{k\ll1}\sim k^{\alpha-1}$ encode the low-energy structure of the system, with the dynamical critical exponent given by $z=\mathrm{min}\{\alpha,\beta\}-1$. In momentum space, the system reduces to a set of independent driven two-level systems (TLSs),
\begin{eqnarray}\label{eq: H_k}
H\!=\!\sum_k\hat\psi^\dagger_k\left[(g-\cos\varphi_k)\tau^z\!+\!\sin\theta_k\tau^x\right]\hat\psi_k
\!=\!\sum_k\!\epsilon_k\gamma^\dagger_k\gamma_k,\quad
\end{eqnarray}
where $\hat\psi_k=(c_k,c^\dagger_{-k})^T$ with $c_k=e^{-i\pi/4}\sum_n e^{ink}c_n$.
The Bogoliubov operators $\gamma_k$ diagonalize the TLS Hamiltonians, yielding quasiparticle energies $\epsilon_k(g)=\sqrt{(g-\cos\varphi_k)^2+\sin^2\theta_k}$ that close at $g_c=\pm\cos\varphi_0$ and vanish as $\epsilon_k(g_c)\sim k^z$ for small momenta.
Crossing this critical point breaks adiabaticity, generating topological defects corresponding to diabatic excitations of the TLSs.
The defect number operator $\hat N=\sum_k\gamma^\dagger_k\gamma_k$ provides a natural measure of nonadiabaticity, allowing a complete statistical characterization of defects through its distribution $P(N)=\langle\delta(\hat N-N)\rangle$ and cumulant generating function $\log\tilde P(\theta)=\log\langle e^{i\theta\hat N}\rangle$, where the expectation value is taken over the time-evolved state.

The exact CD term can also be expressed in terms of the TLS fermionic operators,
\begin{eqnarray}\label{eq: H_1_orthogonal}
    H_1&&=-\frac{\dot g}{2}\sum_k\frac{\sin\theta_k}{(g-\cos\varphi_{k})^2+\sin^2\theta_{k}}\hat\psi^\dagger_k\tau^y\hat\psi_k\nonumber\\
    &&\equiv\sum_kq_k(t)\hat\psi^\dagger_k\tau^y\hat\psi_k,
\end{eqnarray}
where we have introduced the mode-dependent coefficient $q_k(t)$. Applying the Fourier decomposition $q_k(t)=-\frac{1}{\sqrt L}\dot g\sum_m h_m(g)\sin(mk)$, with
$h_m(g)=\frac{1}{\sqrt L}\sum_{k'}\frac{\sin\theta_{k'}\sin(mk')}{(g-\cos\varphi_{k'})^2+\sin^2\theta_{k'}}$,
one obtains an exact representation of the full CD term in an orthogonal operator basis,
$\mathcal O_m=\frac{1}{\sqrt L}\sum_k\sin(mk)\hat\psi^\dagger_k\tau^y\hat\psi_k$,
satisfying $\mathrm{tr}[\mathcal O_l\mathcal O_m]=\delta_{lm}$ (see Methods for details).
This representation is particularly convenient for analyzing locality-controlled truncations, since each operator $\mathcal O_m$ corresponds to pairing terms of range 
$m$, $\mathcal O_m=\frac{1}{\sqrt{2L}}\sum_l(c_l c_{l+m}+c^\dagger_l c^\dagger_{l+m})$.
In interacting spin systems~\cite{Chapman2020characterizationof} that can be mapped to free-fermion forms of Eq.~\eqref{eq: H_0}, these terms translate into multibody spin-string interactions of length $m$.

To this end, we focus on the role of CD in fast quenches, which is particularly relevant for recent experiments~\cite{Visuri2025,Wang2025FastQuenchSimulator} where the dynamics is dominated by the CD term due to the diverging driving rate $\dot g$, rendering the bare Hamiltonian negligible. The time evolution of each $k$-mode then follows
$\partial_t[\psi_{k,1},\psi_{k,2}]^T = -iq^{(n)}_k(t)\tau^y[\psi_{k,1},\psi_{k,2}]^T$,
where the truncated function $q^{(n)}_k(t)=-\dot g\frac{1}{\sqrt L}\sum_{m=1}^n h_m(g)\sin(mk)$ encodes the $n$th-order CD expansion. As the expansion order increases, the dynamics becomes progressively confined to the low-energy sector, corresponding to momenta $k\lesssim n^{-1/(\beta-1)}$ (see Methods). In this regime, the leading-order excitation probabilities are 
\begin{eqnarray}\label{eq: p_k_n}
p_k \approx \cos^2(Cnk^{\beta-1}),
\end{eqnarray}
where $C$ is a non-universal constant, irrelevant for the overall scaling behavior.
As a result, the whole cumulant generating function together with all defect number cumulants will inherit this behavior, resulting in a universal scaling law,
\begin{eqnarray}\label{eq: kappa_n}
    \log\tilde P^{(n)}(\theta;0)&=&\sum_k\log\left[1+(e^{i\theta}-1)p_k\right]\propto n^{-1/(\beta-1)},\,\quad\\
    \kappa^{(n)}_q(0)&=&-\partial^q_{i\theta}\log\tilde P^{(n)}(\theta;0)\Big\vert_{\theta=0}\propto n^{-1/(\beta-1)}.
\end{eqnarray}
Thus, in the large-$n$ limit the defect number distribution approaches a Gaussian form with mean $\kappa^{(n)}_1(0)$ and variance $\kappa^{(n)}_2(0)$, i.e., $P^{(n)}(N)\sim\mathcal{N}[\kappa^{(n)}_1(0),\kappa^{(n)}_2(0)]$.
In Methods, we show that a complementary local expansion scheme achieved by the decomposition $q^{(n)}_k=\sum_{m=1}^n\tilde h_m(g)\sin(m\theta_k)$, which uses a non-orthogonal operator basis, leads to the same universal scaling. This further reinforces the universal conjecture on the effect of local CD in fast quenches.

These universal characteristics also offer an ideal setting to examine the effects of local CD in the slow-driving limit. For linear protocols $g(t)=-t/T,\quad t\in[-\infty,0]$, and for $k\ll n^{-1/(\beta-1)}$ with time scales $T\gg n^2$, adiabaticity is violated by the bare Hamiltonian, $Tk^{2z}\sim O(1)$, while the $n$th-order CD contribution gets strongly suppressed, $q^{(n)}_k\sim nk^{-(\beta-1)}/T\sim nk^{(\beta-1)}/(Tk^{2(\beta-1)})\ll1$. Hence, this regime is effectively insensitive to CD, and defect generation follows the KZ scaling law. For larger momenta $k\gg n^{-1/(\beta-1)}$, adiabaticity is trivially maintained by the effects of CD, as in fast quenches.
For $T\ll n^2$, modes with $k\lesssim n^{-1/(\beta-1)}$ exhibit a fast-quench plateau similar to that of the bare dynamics~\cite{Zeng2023Universal}, while higher-energy modes are adiabatically driven by CD. Thus, the sudden-quench regime extends up to $T\propto n^2$, defining an effective low-energy CD fast-quench breakdown scale 
\begin{eqnarray}\label{eq: CD_fastquenchscale}
T^\mathrm{CD}_\mathrm{fast}\propto n^2.
\end{eqnarray}

\subsection*{Exact defect statistics in the TFIM}

As a first exact benchmark of the general framework above, we consider the transverse-field Ising model (TFIM), an archetypal setting for quantum critical dynamics~\cite{Zurek2005Dynamics, Dziarmaga2005Dynamics, Sachdev2011Quantum, Suzuki2012Quantum}
\begin{eqnarray}\label{eq:H_TFIM}
\hat{H}(t) = - \sum_{j=1}^L \left( \hat{\sigma}^z_j \hat{\sigma}^z_{j+1} + g(t) \hat{\sigma}_j^x \right),
\end{eqnarray}
with a linear ramp $g(t)=g(0)(1-t/T),\,g(0)\gg1$. Applying the Jordan-Wigner and a Fourier transformation~\cite{Suzuki2012Quantum}, one arrives at the same TLS Hamiltonian in Eq.~\eqref{eq: H_k} with $\varphi_k=\theta_k=k$ and $z=1$.

\begin{figure*}[b]
    \centering
    \includegraphics[width=\linewidth]{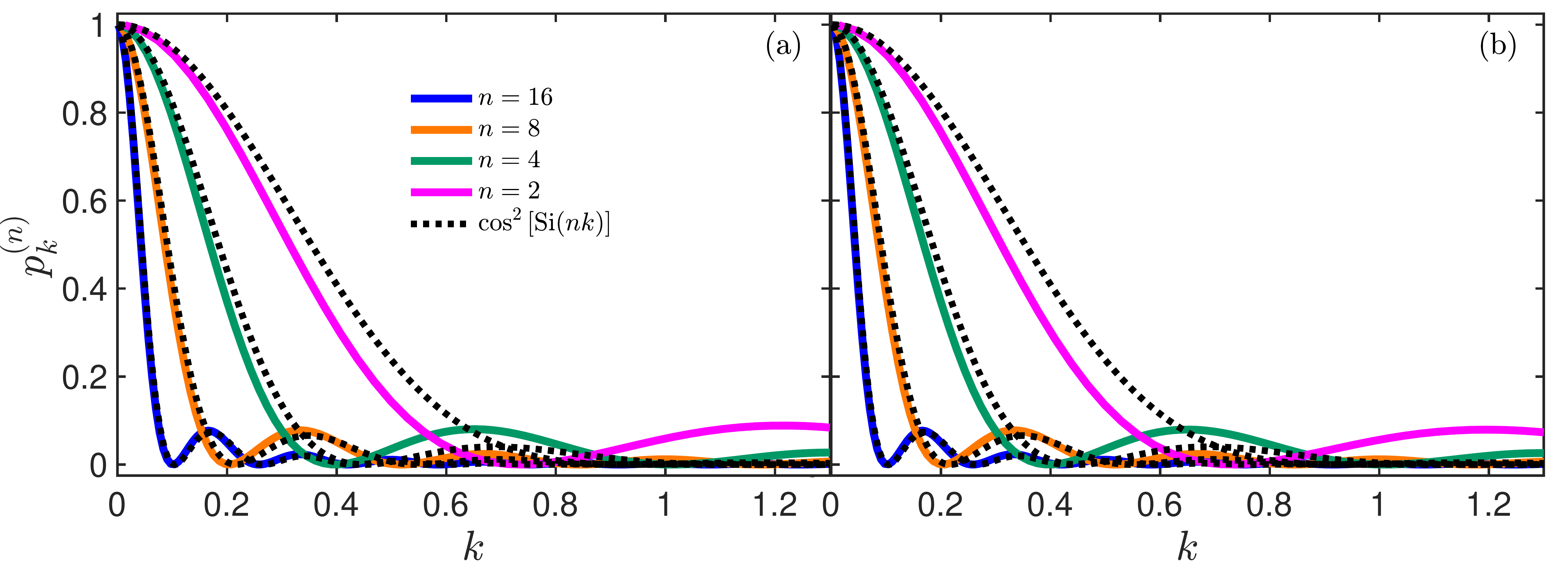}\\    
    \caption{Fast-quench excitation probabilities for several CD expansion orders, $n=2,4,8,16$. The numerical results agree well with the limiting form in Eq.~\eqref{eq:p_k_tfim}, both deep in the sudden-quench regime at $T=10^{-6}$, $(a)$, and close to the crossover regime at $T=4$, $(b)$. Even for the lowest order, $n=2$, the approximate form of $p^{(n)}_k$ captures well the dominant low-momentum behavior ($L=1600$). }
    \label{fig:p_k_fast}
\end{figure*}

Simultaneous excitations in modes $(-k,k)$ correspond to kink pairs due to the $\mathbb Z_2$ symmetry of the ferromagnetic ground state. They are naturally counted by the kink-number operator, $\hat N=\frac12\sum_j (1-\hat\sigma^z_j\hat\sigma^z_{j+1})$. For slow ramps across the critical point, $g_c=\pm1$, excitation probabilities are captured by the Landau-Zener (LZ) probabilities, $p_k = \langle \gamma^\dagger_k \gamma_k\rangle \approx e^{-2\pi k^2 T}$, yielding $\langle N(T)\rangle \approx L/\sqrt{8\pi^2 T}$, which is in agreement with the original argument of the KZ mechanism predicting $\langle\hat N\rangle\sim T^{-\frac{d\nu}{z\nu+1}}$ for point-like defects in $d$ dimensions with $z=\nu=1$ in $d=1$ for the TFIM~\cite{Damski05,DamskiZurek06,Dziarmaga2005Dynamics}. Defect number cumulants of higher order follow the same universal power-law decay as the average, $\kappa_q\propto \langle \hat N(T)\rangle$, while fast quenches break the KZ scaling~\cite{Zeng2023Universal,Grabarits2025}, and in the sudden limit $p_k \to \cos^2(k/2)$, giving $\langle N(T)\rangle \approx L/2$.

 In the TFIM, the general orthogonal operator subspace construction is identical to the Krylov expansion~\cite{delcampo12,Damski2014Counterdiabatic,Takahashi2024Shortcuts},
\begin{eqnarray}
H_1 &=& -\dot g \frac{2}{L}\sum_{m=1}^{L/2}\sum_{k'}\frac{\sin k'\sin(mk')}{1+g^2-2g\cos k'}\sum_{k>0}\sin(m k) \hat\psi_k^\dagger \tau^y_k \hat\psi_k\nonumber\\
&\equiv&\sum_{k>0} q_k(t) \hat\psi_k^\dagger \tau^y_k \hat\psi_k,
\end{eqnarray}
for which the $n$-th order local expansion is obtained by the truncation~\cite{Damski2014Counterdiabatic}
\begin{eqnarray}
q^{(n)}_k(g) =  -\dot g\sum_{m=1}^n \frac{\sin(k m)}{2} \frac{g^{2m}+g^L}{g^{m+1}(1+g^L)}.
\end{eqnarray}

As in the general setting, the fast-quench dynamics is governed entirely by the CD term. For the TFIM, the exact \(n\)-th order excitation probability can be written as $p^{(n)}_k
=
\sin^2\frac{k}{2}\,\sin^2[{\rm Si}(n,k)]
+\cos^2\frac{k}{2}\,\cos^2[{\rm Si}(n,k)]
-\frac12 \sin k\,\sin[2{\rm Si}(n,k)]$~\cite{supp},
where
${\rm Si}(n,k)=\sum_{m=1}^n \frac{\sin(km)}{m}$.
(For further exact results of the time-evolved wave-function, see~\cite{supp}.)

\begin{figure}
\centering
\includegraphics[width=0.49\linewidth]{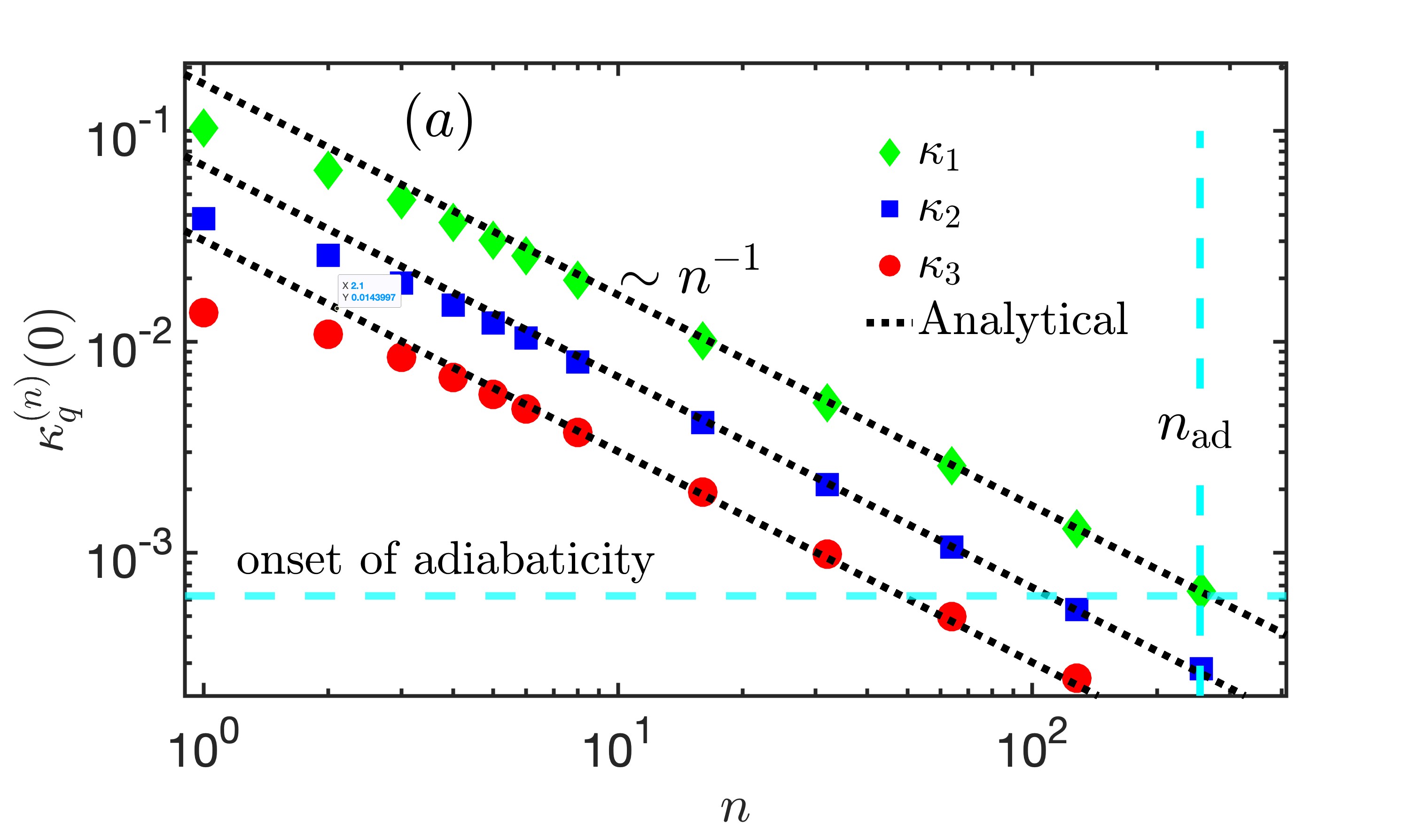}
\includegraphics[width=0.49\linewidth,trim={0 0 0 2cm}, clip]{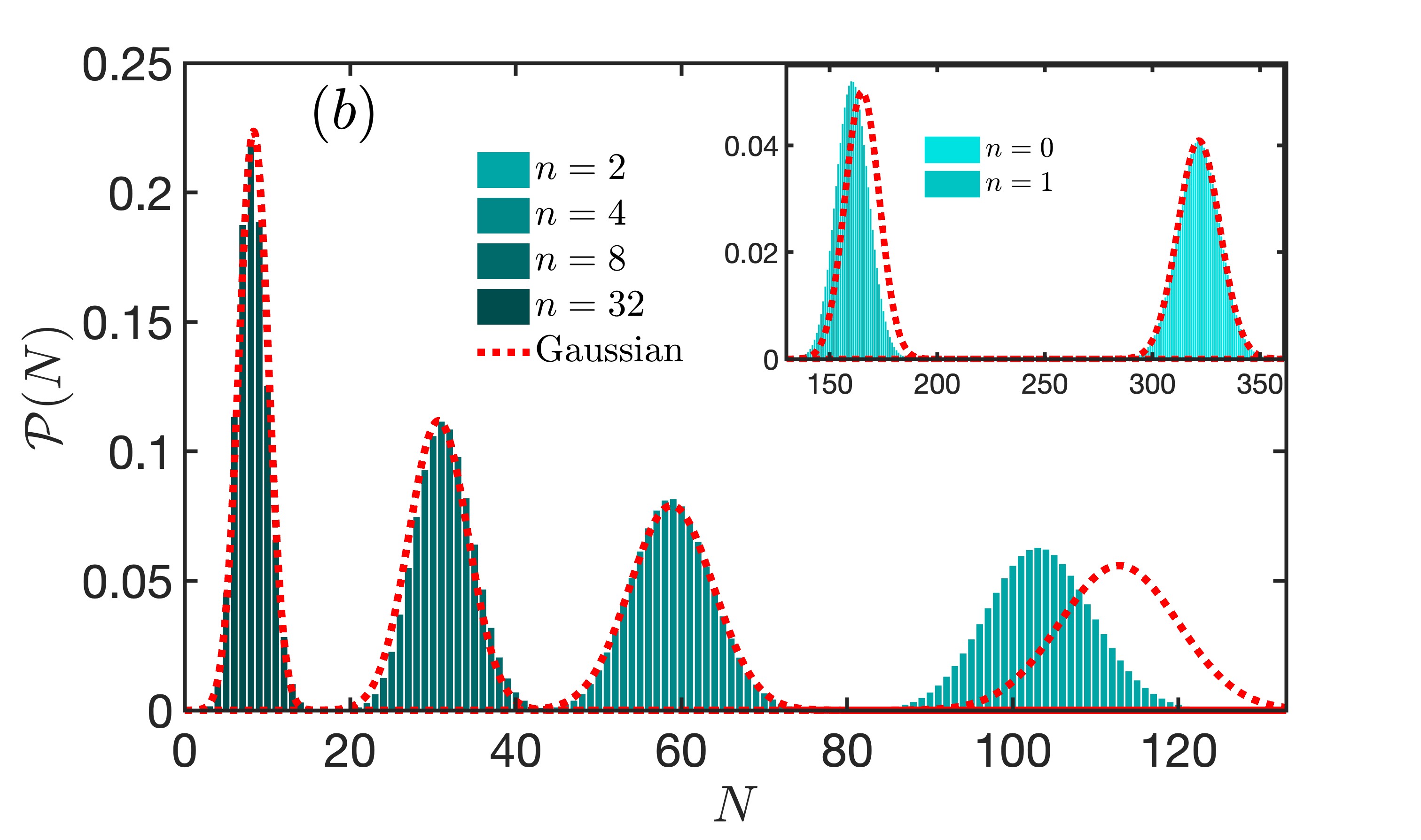}
\caption{$(a)$ Universal scaling of the first three defect cumulants with CD order in the TFIM under fast driving. The analytical slopes are in excellent agreement with the numerical data ($L=1600$, $T=1$).
$(b)$ Defect-number distributions in the fast-quench regime at $T=2$ for several values of $n$, consistent with the large-$n$ Gaussian prediction. As the CD order is increased, defect suppression manifests in the gradual shift of the distributions towards zero and in their progressive narrowing. Inset: defect statistics without CD and for $n=1$, together with the exact low-order result ($L=1600$).}
\label{fig:KinkStat_sudden}
\end{figure}

For small momenta, this exact expression reduces to the universal leading-order form of Eq.~\eqref{eq: p_k_n},
\begin{eqnarray}\label{eq:p_k_tfim}
    p_k^{(n)}\approx \cos^2[\mathrm{Si}(nk)]\approx \cos^2(nk),
\end{eqnarray}
consistent with the scaling variable \(nk\) for \(z=1\). Figure~\ref{fig:p_k_fast} numerically verifies this low-momentum form. The agreement remains excellent both deep in the sudden-quench limit, 
$T=10^{-6}$, and even for moderate driving times, 
$T=4$. Additional results, including the scaling collapse of the excitation probabilities and exact time-evolved wave-function components, are provided in~\cite{supp} in Supplementary Fig.~S1 and Supplementary Fig.~S2, respectively.

This asymptotic form also allows one to extend the momentum integrals up to $\infty$, leading to the cumulant generating function
\begin{equation}\label{eq:fast_logP}
    \log\tilde P^{(n)}(\theta;0)\approx\frac{1}{n}\int_0^\infty\!\!\mathrm dx\log\left\{1+(e^{2i\theta}-1)\cos^2\left[\mathrm{Si}(x)\right]\right\}.
\end{equation}
As a result, the cumulants satisfy \(\kappa_q^{(n)}(0)\propto 1/n\), with explicit sub-Poissonian ratios \(\kappa_2/\kappa_1\approx0.82\) and \(\kappa_3/\kappa_1\approx0.76\).

Remarkably, already the first-order CD correction strongly modifies the fast-quench cumulants relative to the uncontrolled case:
\(\kappa^{(1)}_1(0)\approx0.2\,L\),
\(\kappa^{(1)}_2(0)\approx0.096\,L\),
and
\(\kappa^{(1)}_3(0)\approx0.112\,L\),
to be contrasted with
\(\kappa_1(0)=L/2\),
\(\kappa_2(0)=L/4\),
and
\(\kappa_3(0)=0\)
without CD. The numerical verification of the CD-order scaling of the first three cumulants is shown in Fig.~\figpanel{fig:KinkStat_sudden}{a}. 
As shown in~\cite{supp}, the scaling of the mean also allows one to extract the adiabatic threshold of the expansion order, $n^\mathrm{CD}_\mathrm{ad}\approx\frac{1.05}{2\pi}L$, which is likewise indicated in Fig.~\figpanel{fig:KinkStat_sudden}{a}.
Further results on the crossover between the fast-quench regime and the slow-driving limit for different CD expansion orders are presented in Methods, while additional exact results for the time-evolved wave function are given in~\cite{supp}.

We also show the effect of the local expansion on the full defect distribution. As each mode excitation follows an independent Bernoulli process, the defect distribution approaches a normal form determined by its first two cumulants. Consistent with their scaling, both the mean and the variance vanish with increasing $n$, as shown in Fig.~\figpanel{fig:KinkStat_sudden}{b}. Remarkably, even for $n=2$, only small deviations are found with respect to the large $n$ limit.

\begin{figure*}
    \centering
    \includegraphics[width=0.6\linewidth]{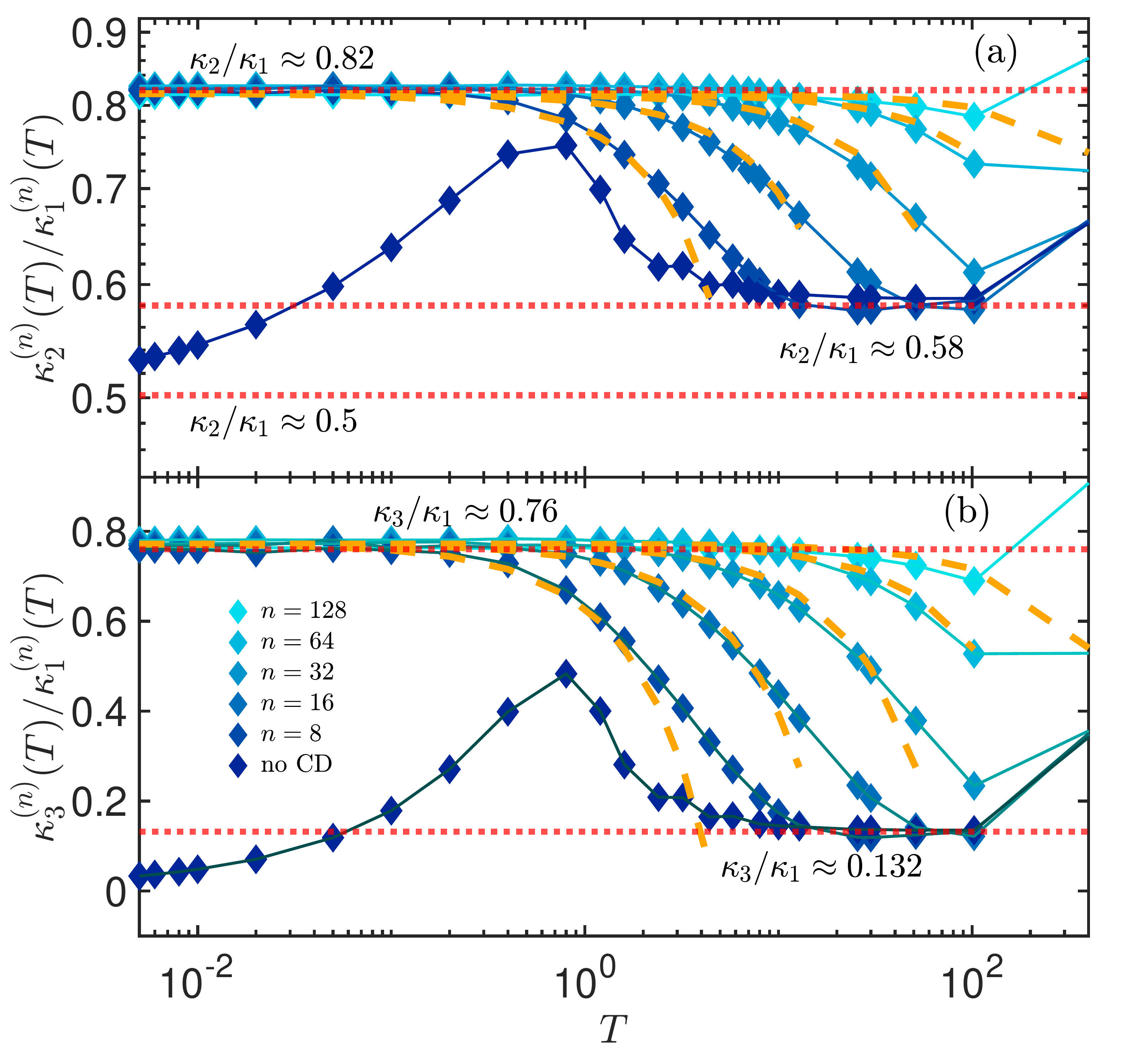}    
    \caption{Cumulant ratios as functions of the driving time for several CD expansion orders, $n=0,8,16,32,64,128$. $(a)$ The variance ratio $\kappa_2/\kappa_1$ shows a smooth crossover between the fast-quench plateaus with and without CD and the CD-independent KZ scaling limit. $(b)$ The skewness ratio $\kappa_3/\kappa_1$ shows an analogous crossover, with the no-CD case converging to zero in the fast-quench regime. Orange dashed lines indicate the leading-order expansion around the sudden limit.}
    \label{fig:cumulant_ratios}
\end{figure*}

Finally, we demonstrate that the cumulant ratios capture the crossover between the fast- and slow-driving regimes for different CD orders. As shown in Fig.~\figpanel{fig:cumulant_ratios}{a,b}, \(\kappa_2/\kappa_1\) and \(\kappa_3/\kappa_1\) interpolate between constant fast-quench and KZ plateaus. In the fast-quench regime, the large-\(n\) values are \(\kappa_2/\kappa_1\approx0.82\) and \(\kappa_3/\kappa_1\approx0.76\), compared to \(\kappa_2/\kappa_1\approx0.5\) and \(\kappa_3/\kappa_1=0\) without CD. For \(T\gg n^2\), the ratios approach the KZ limit when \(n<n^\mathrm{CD}_\mathrm{ad}\), with \(\kappa_3/\kappa_1\approx0.132\).

\subsection*{Demonstration in the long-range Kitaev models}

As a further test of the universality of the fast-quench defect generation under local CD, we analyze the long-range Kitaev model (LRKM), describing spinless fermions hopping on a one-dimensional lattice in the presence of $p$-wave pairing.
The LRKM can conveniently be formulated in momentum space as
\begin{eqnarray}\label{eq: LRKM_Ham}
    H(t)&=&-2g(t) \sum_{i=1}^L c^\dagger_i c_i -\sum_{i=1}^L \sum_{r=1}^{L/2} \left[ j_{r,\alpha} c^\dagger_ic_{i+r} + d_{r,\beta} c^\dagger_{i} c^\dagger_{i+r} +\,\mathrm{h.c.} \right],\\
    H&=&2 \sum_{k>0} \hat{\psi}_k^\dagger \left[ [g(t)-j_\alpha(k)]\tau^z + d_\beta(k) \, \tau^x \right] \hat{\psi}_k=\sum_k\hat\psi^\dagger_k\left[(g-\cos\varphi_k)\tau^z+\sin\theta_k\tau^x\right]\hat\psi_k\nonumber\\
    &\equiv&2 \sum_{k>0} \hat{\psi}_k^\dagger H_k(t) \hat{\psi}_k,\\
    j_{\alpha,r}&=&\mathcal N^{-1}_\alpha r^{-\alpha},\quad j_\alpha(k)=\mathcal N^{-1}_\alpha\sum_{r=1}^{L/2}\,r^{-\alpha}\,\cos(kr)\approx \mathrm{Re}\left[\mathrm{Li}_\alpha(e^{ik})\right]\propto \cos\varphi_k,\\
    d_{\beta,r}&=&\mathcal N^{-1}_\beta r^{-\beta},\quad d_\beta(k)=\mathcal N^{-1}_\beta\sum_{r=1}^{L/2}\,r^{-\beta}\,\sin(kr)\approx \mathrm{Im}\left[\mathrm{Li}_\beta(e^{ik})\right]\propto\sin\theta_k,
\end{eqnarray}
with $\mathcal N_\gamma=2\sum_{r=1}^{L/2}r^{-\gamma}\approx 2\zeta(\gamma)$
and where $\hat\psi_k=(c_k,c^\dagger_{-k})^T$ contains the creation and annihilation operators of the $k$-th mode. Thus, the LRKM realizes the general two-level structure with tunable short- and long-range critical behavior.
The hopping and pairing amplitudes decay algebraically, 
$j_r=\mathcal N^{-1}_\alpha\, r^{-\alpha}$ and $\Delta_r=\mathcal N^{-1}_\beta\, r^{-\beta}$, 
with exponents $\alpha,\beta>1$ and normalization factors $\mathcal N_{\alpha,\beta}=2\sum_{r=1}^{L/2}r^{-\alpha,\,-\beta}$ ensuring extensivity.
\begin{figure}[b]
\centering
\includegraphics[width=0.49\linewidth]{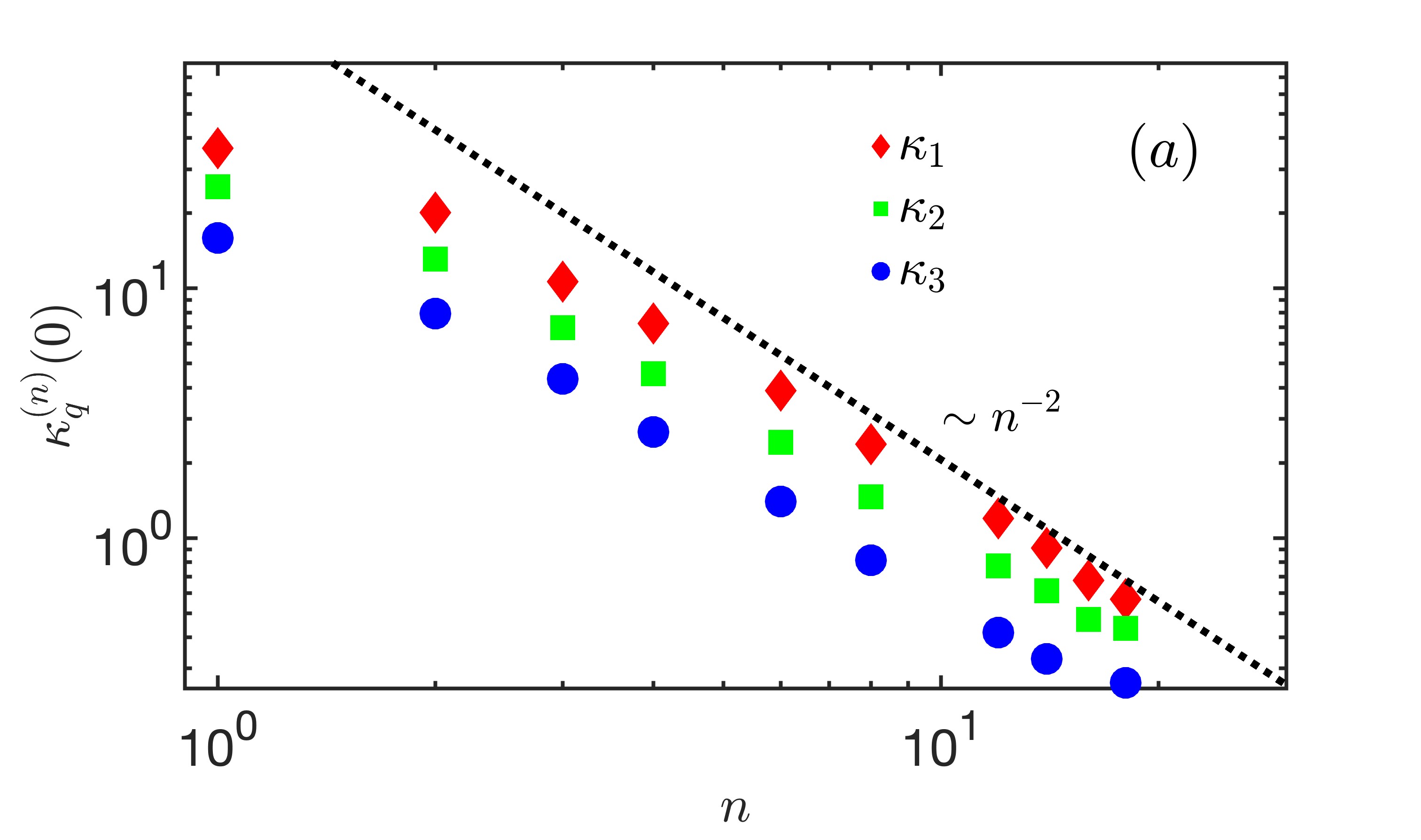}
    \includegraphics[width=0.49\linewidth,trim={0 0 0 1.5cm}]{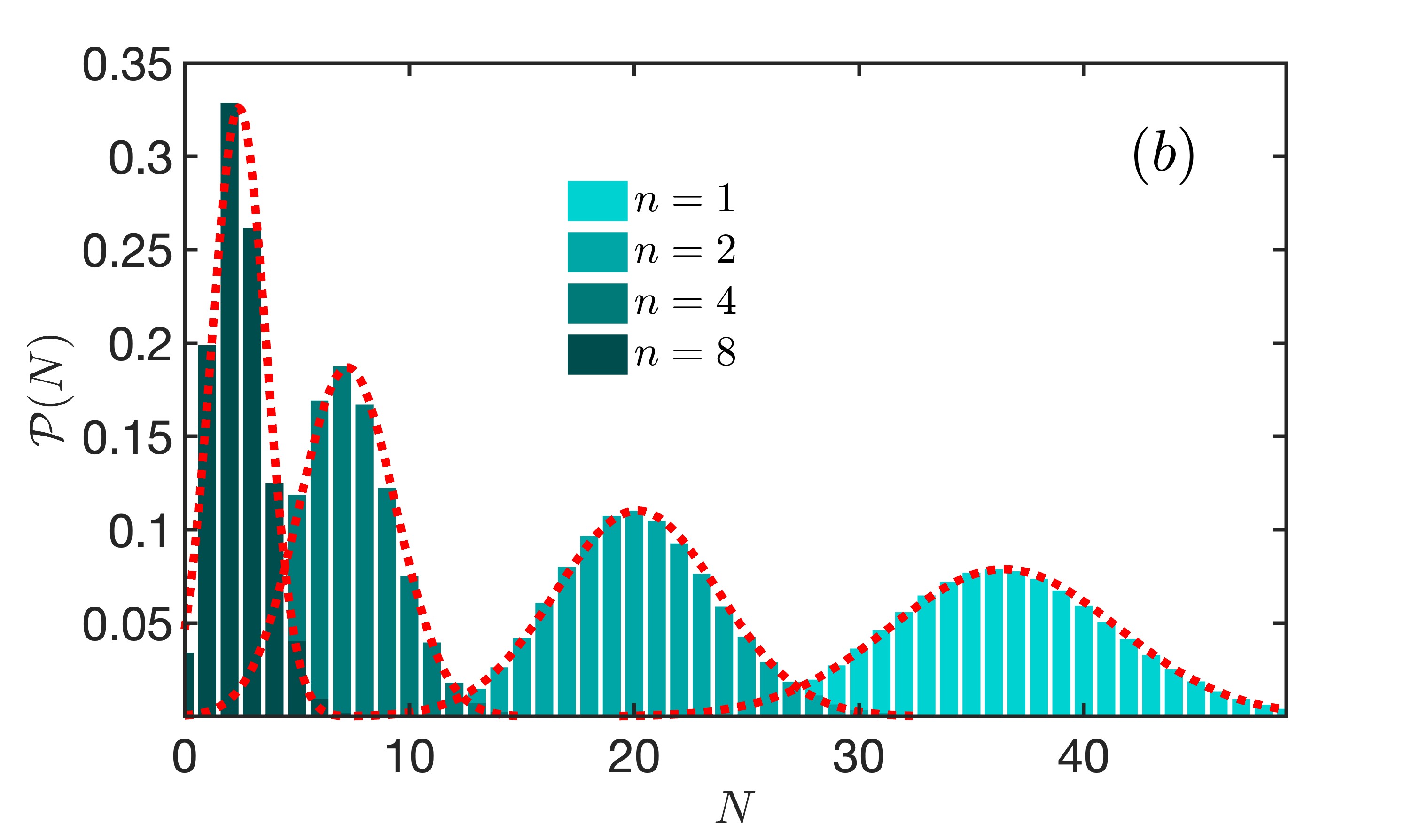}
    \caption{$(a)$ Fast-quench defect cumulants in the LRKM as a function of CD locality, obtained from the orthogonal expansion, Eq.~\eqref{eq: orth_expansion}, for long-range hopping and pairing ($\alpha=2.5$, $\beta=1.5$). The first three cumulants exhibit the predicted power-law decay, $\propto n^{-2}$ ($L=800$).
$(b)$ Defect-number distributions in the LRKM for various CD expansion orders, approaching a Gaussian limit. As the order grows, the histograms exhibit a rapid convergence towards zero while becoming progressively narrower.}
    \label{fig:LRKM_Stat}
\end{figure}
With these exponents, the LRKM exhibits a second-order QPT at $g_c=2$~\cite{Kitaev_Majorona_LR_2001}, which is robust against variations of the long-range exponents $\alpha,\beta>1$~\cite{Vodola_LRK, Dutta_LRK, LongRangePowerLawSC_Delgado_2017, LRK_powerlawDelAnna2017, Defenu_LRKM}. The bulk topological invariant $w$ counts the number of doubly degenerate Majorana zero modes (MZMs): the trivial phase $g>g_c$ has $w=0$, while for $g<g_c$ two MZMs localize exponentially at the chain ends under open boundaries, $w=1$~\cite{Albrecht2016MZM_edgestates, MZMmodes_Science_20212}.
 At low momenta, the momentum space representation after Fourier decomposing the fermionic operators yields $\cos\varphi_k=\mathrm{Re}\left\{\mathrm{Li}_\alpha(e^{ik})\right\}= 1-C_\alpha k^{\mathrm{min}\{2,\alpha-1\}}+O(k^{\mathrm{min}\{\alpha,3\}}),\quad\sin\theta_k=\mathrm{Im}\left\{\mathrm{Li}_\beta(e^{ik})\right\}\sim k^\mathrm{min\{1,\beta-1\}}+O(k^{\mathrm{min}\{\beta,2\}})$ with $z=\mathrm{min}\{\alpha,\beta\}-1$. This makes the LRKM particularly well suited to test the central prediction of our framework: namely, that the fast-quench locality scaling of the defect cumulants is controlled by the low-momentum behavior of $\theta_k$, and therefore by 
$\beta$, rather than in general by the dynamical critical exponent $z$.

 First, we focus on the long-range regime with dominating hopping terms, $\alpha\leq \mathrm{min}\{2,\beta\}$, where the low-momentum energy dispersion relation is governed by $\cos\varphi_k,\,z=\alpha-1$, giving rise to nontrivial long-range effects but with the defect formation still respecting the KZ scaling law for slow drivings~\cite{Defenu_LRKM}. As shown in Fig.~\figpanel{fig:LRKM_Stat}{a}, numerical results from the exact orthogonal expansion scheme, Eq.~\eqref{eq: H_1_orthogonal}, are in excellent agreement with the universal scaling predictions. In particular, the first three cumulants precisely follow the same universal power-law of the CD locality, $\kappa_q\propto n^{-1/(\beta-1)}$, demonstrating explicitly that the scaling is governed by the dispersion of $\theta_k$ rather than by the dynamical critical exponent. The limiting Gaussian behavior is further demonstrated in Fig.~\figpanel{fig:LRKM_Stat}{b}, showing trends analogous to those of the TFIM, with histograms progressively narrowing and shifting towards zero as the CD order increases.

 \begin{figure}[ht!]
  \centering
  \includegraphics[width=0.495\textwidth,trim={2cm 0 10cm 4cm},clip]{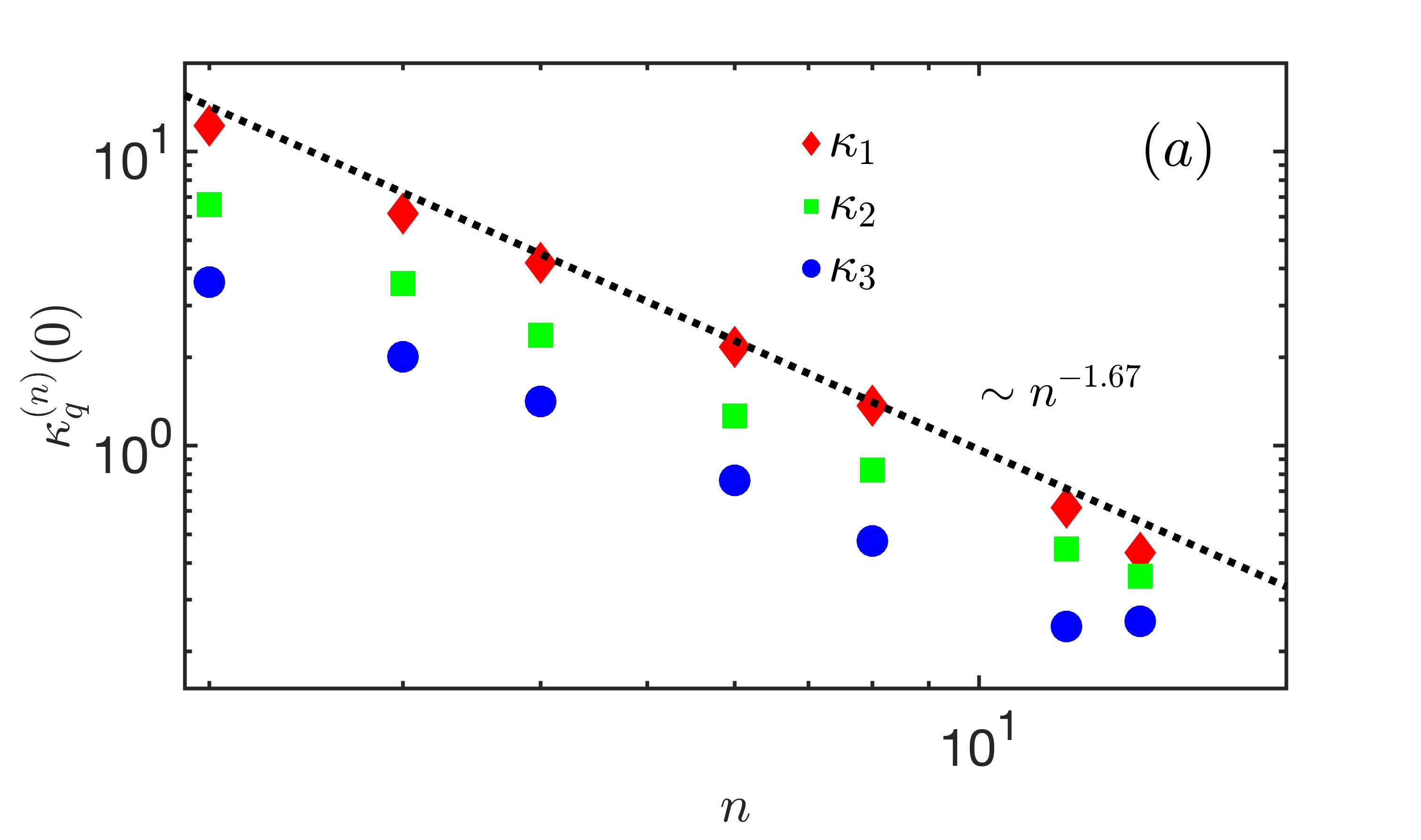}\hfill
  \includegraphics[width=0.465\textwidth,trim={7cm 0 10cm 0},clip]{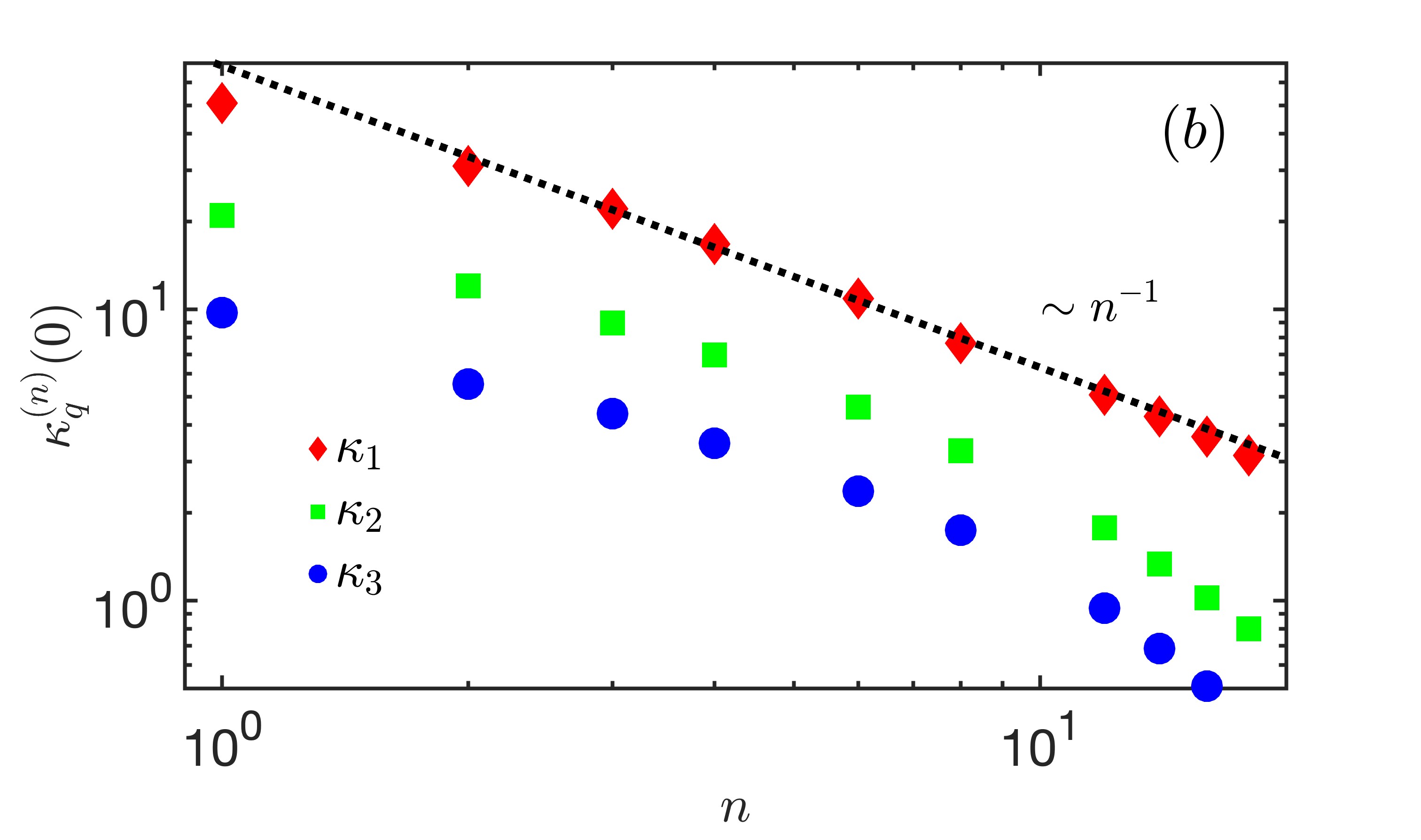}\hfill
  \caption{Fast-quench locality scaling of the defect cumulants in the LRKM, obtained with the Krylov expansion, for $(a)$ $\alpha=1.4,\,\beta=1.6$ in the dynamical-scaling long-range regime and $(b)$ $\beta=5,\,\alpha=1.2$ in the short-range pairing and long-range hopping regime. In all cases, the first three cumulants decay proportionally according to the predicted power law $n^{-1/(\beta-1)}$.}
  \label{fig:LRKM_fast}
\end{figure}
 
Finally, we also show the cumulant scalings for all the other regimes of both short- and long-range hopping and pairing interactions. As shown in Fig.~\figpanel{fig:LRKM_fast}{a}, for long-range pairing and hopping terms in the dynamical scaling regime~\cite{Defenu_LRKM}, $\alpha=1.4,\,\beta=1.6$ the expected $\kappa_q\propto n^{-1.67}$ is observed. By contrast, for short-range pairings and long-range hoppings in Fig.~\figpanel{fig:LRKM_fast}{b} for $\beta=5,\,\alpha=1.2$, $\kappa_q\propto n^{-1}$ is found with high accuracy, in both cases with the numerical implementation of the Krylov expansion scheme. 
In the Methods, we further present numerical results based on an alternative, non-orthogonal operator expansion in the $\beta=5,\,\alpha=5$ short-range regime.
Taken together, these results show that the universal fast-quench scaling with CD locality persists across all regimes considered, independently of whether the model is short- or long-ranged and independently of which expansion scheme is used.

Further analytical and numerical results demonstrating the universal effect of CD in slow processes and the CD fast-quench breakdown scale are presented in~\cite{supp} in Supplementary Fig.~S3.

\section*{Discussion}

Counterdiabatic driving provides a powerful method to suppress excitations in driven systems, yet its performance under approximate, locally tailored protocols has remained elusive. We develop a novel and analytically tractable local CD expansion scheme. Within this framework, we establish a universal scaling theory of defect generation in both the fast- and slow-driving regimes. In the fast-quench limit, all cumulants scale universally with the inverse of the CD expansion order, with their limiting statistics converging to a Gaussian distribution. The CD-fast-quench breakdown grows quadratically with the expansion order, beyond which defect generation becomes insensitive to CD and follows the conventional KZ scaling. These predictions are confirmed in exactly solvable models.
Our analytical findings thus provide a general framework for assessing the efficiency of local CD protocols, with broad implications for quantum simulation, quantum control, quantum thermodynamics, and counterdiabatic quantum algorithms.

\section*{Methods}
\subsection*{Exact orthogonal CD expansion for arbitrary exactly solvable critical systems}

In this section, we provide the analytical details of the local CD expansion in an orthogonal operator basis and derive the universal fast-quench scaling of the defect cumulants discussed in the main text. We consider the general quadratic Hamiltonian
\begin{eqnarray}
H(t)=\sum_k\hat\psi^\dagger_k H_k(t)\hat\psi_k,
\end{eqnarray}
with low-energy structure
\begin{eqnarray}
H_k(t)=(g-\cos\varphi_k)\tau^z+\sin\theta_k\tau^x,
\qquad
\varphi_k\sim k^{\alpha-1},\quad
\theta_k\sim k^{\beta-1},
\end{eqnarray}
so that the energy gap closes as \(\epsilon_k\sim k^z\) with \(z=\min\{\alpha,\beta\}-1\). In real space, this corresponds to
\begin{eqnarray}
H=\sum_{r,l}\left(j_r c_l c^\dagger_{l+r}+d_r c_l c_{l+r}+\mathrm{h.c.}\right),
\end{eqnarray}
where the hopping and pairing terms determine \(\cos\varphi_k\) and \(\sin\theta_k\), respectively.

The exact CD term is then
\begin{eqnarray}
H_1=-\dot g\sum_k\frac{\sin\theta_k}{2\left[(g-\cos\varphi_k)^2+\sin^2\theta_k\right]}
\hat\psi^\dagger_k\tau^y\hat\psi_k.
\end{eqnarray}
We expand it in the orthogonal basis
\begin{eqnarray}
H^{(n)}_1=\sum_k q^{(n)}_k(t)\hat\psi^\dagger_k\tau^y\hat\psi_k,
\end{eqnarray}
with
\begin{eqnarray}\label{eq: orth_expansion}
q^{(n)}_k(t)=-\dot g\frac{1}{\sqrt L}\sum_{m=1}^n h_m(g)\sin(mk),
\qquad
h_m(g)=\frac{1}{\sqrt L}\sum_k
\frac{\sin\theta_k\sin(mk)}{(g-\cos\varphi_k)^2+\sin^2\theta_k}.
\end{eqnarray}
Although this construction is not identical to the Krylov expansion, its \(n\)-th order terms form an orthogonal set of operators,
\begin{eqnarray}
H^{(n)}_1&=&-\frac{\dot g}{\sqrt L}\sum_{m=1}^n h_m(g)\mathcal O_m,\\
\mathcal O_m&=&\frac{1}{\sqrt L}\sum_k\sin(mk)\hat\psi^\dagger_k\tau^y\hat\psi_k,
\qquad
\tr[\mathcal O_n\mathcal O_m]=\delta_{nm}.
\end{eqnarray}
In real space, this yields a locality-controlled truncation of the full CD term,
\begin{eqnarray}
H^{(n)}_1=-\frac{\dot g}{\sqrt{2L}}\sum_{m=1}^n h_m(g)\sum_l
\left(c_lc_{l+m}+c^\dagger_l c^\dagger_{l+m}\right).
\end{eqnarray}

We now focus on the fast-quench regime, where the CD term dominates the dynamics. The time-dependent Schr\"odinger equation for each mode becomes
\begin{eqnarray}
\partial_t
\begin{bmatrix}
\psi_{k,1}(t)\\
\psi_{k,2}(t)
\end{bmatrix}
=
-i q^{(n)}_k(t)\tau^y
\begin{bmatrix}
\psi_{k,1}(t)\\
\psi_{k,2}(t)
\end{bmatrix},
\end{eqnarray}
with solution
\begin{eqnarray}
\begin{bmatrix}
\psi_{k,1}(t)\\
\psi_{k,2}(t)
\end{bmatrix}
=
e^{-i\int_0^T dt\, q^{(n)}_k(t)\tau^y}
\begin{bmatrix}
0\\
1
\end{bmatrix}.
\end{eqnarray}
Thus, the excitation probability is determined by the phase integral \(\int_0^T dt\, q^{(n)}_k(t)\). Using Eq.~\eqref{eq: orth_expansion}, one finds
\begin{eqnarray}\label{eq: varphi_k}
\int_0^T dt\,q^{(n)}_k(t)
&=&
\frac{1}{L}\int_0^{-\infty}dg
\sum_{k'}\sum_{m=1}^n
\frac{\sin\theta_{k'}\sin(mk')\sin(mk)}
{(g-\cos\varphi_{k'})^2+\sin^2\theta_{k'}}\nonumber\\
&\sim&
\frac{1}{L}\sum_{m=1}^n\sum_{k'}
\sin\theta_{k'}\sin(mk')\sin(mk)
\frac{\mathrm{arccot}[\cos\varphi_{k'}/\sin\theta_{k'}]}{\sin\theta_{k'}}\nonumber\\
&\sim&
C\,n\,k^{\beta-1},
\end{eqnarray}
where \(C\) is a non-universal constant. Therefore, in the low-momentum regime,
\begin{eqnarray}
p_k
=
\left|
\cos\frac{\theta_k}{2}\cos\!\left(Cnk^{\beta-1}\right)
-
\sin\frac{\theta_k}{2}\sin\!\left(Cnk^{\beta-1}\right)
\right|^2
\approx
\cos^2\!\left(Cnk^{\beta-1}\right).
\end{eqnarray}

The cumulant generating function then takes the asymptotic form
\begin{eqnarray}
\log\tilde P^{(n)}(\theta;0)
&=&
\sum_k\log\left[1+(e^{i\theta}-1)p_k\right]\nonumber\\
&\approx&
\frac{L}{2\pi}\int_0^\pi dk\,
\log\left[1+(e^{i\theta}-1)p_k\right]\nonumber\\
&\propto&
n^{-1/(\beta-1)}
\int_0^{x_{\mathrm{max}}}dx\,
\log\left[1+(e^{i\theta}-1)\cos^2(x)\right],
\end{eqnarray}
where \(x=Cnk^{\beta-1}\) and \(x_{\mathrm{max}}=Cnk_{\mathrm{max}}^{\beta-1}\). As a result, all defect cumulants inherit the same universal scaling,
\begin{eqnarray}
\kappa_q\propto n^{-1/(\beta-1)},
\qquad
q=1,2,3,\dots
\end{eqnarray}
This shows that the CD expansion scaling of the fast-quench defect statistics is controlled by the low-momentum behavior of \(\theta_k\), namely by \(\beta\), whereas the dynamical critical exponent \(z\) governs the closing of the energy gap.

\subsection*{Alternative CD expansion in non-orthogonal operator basis}
Next, we show that an alternative local expansion scheme, but without orthogonal operators, achieves the same leading order behavior. In particular, we employ a different generalization of the known results for the TFIM~\cite {Damski2014Counterdiabatic,delcampo12} for
arbitrary $\theta_k$. We generalize the Krylov expansion for TFIM by replacing $k$ by $\theta_k$,
\begin{eqnarray}\label{eq: nonorth_expansion}
q_k(t)=-\dot g\frac{1}{\sqrt L}\sum_mh_m(g)\sin(m\theta_k).
\end{eqnarray}
\begin{figure*}[b]
    \centering
    \includegraphics[width=0.5\textwidth,trim={7cm 0 10cm 0},clip
  ]{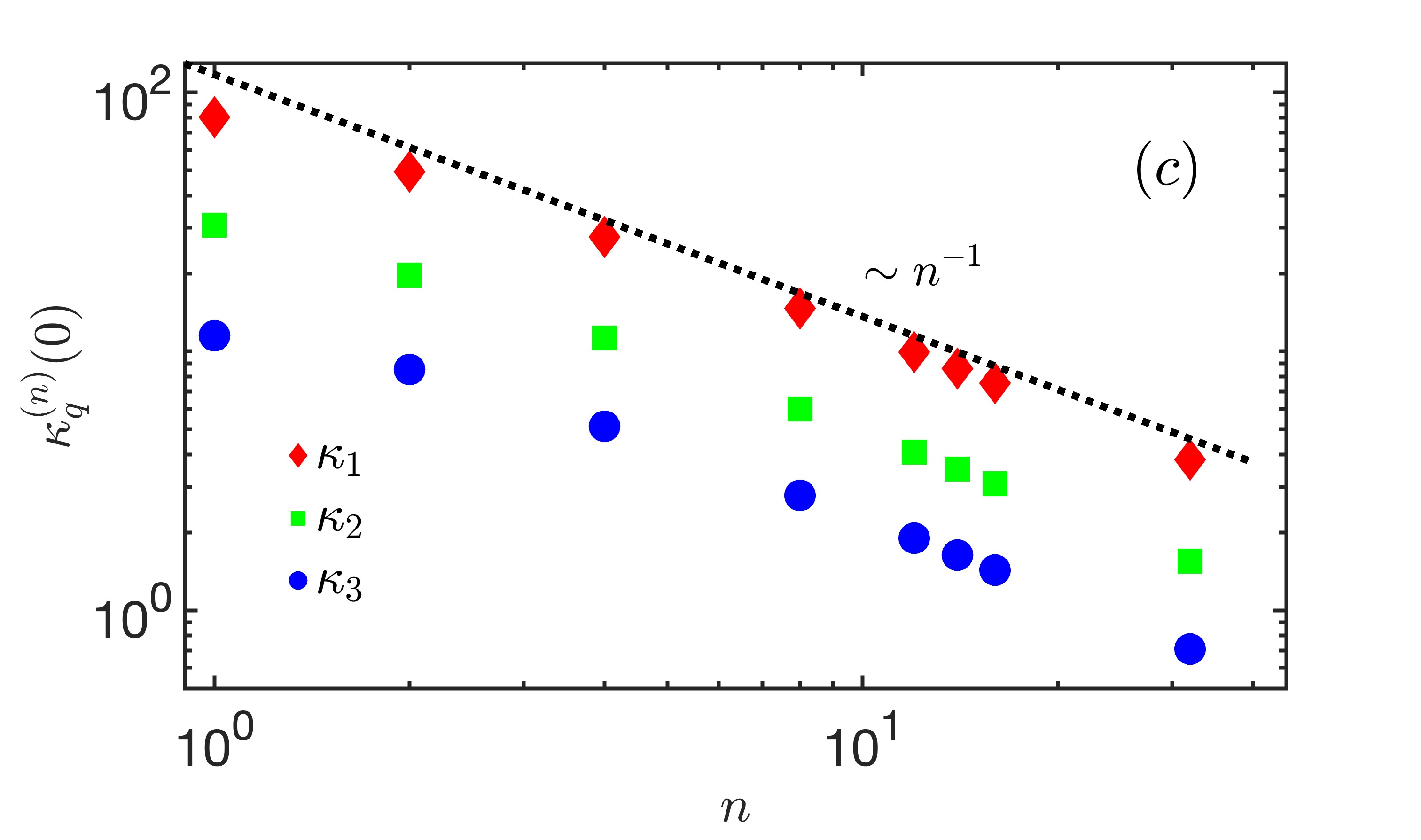}
    \caption{Fast-quench locality scaling of the defect cumulants in the LRKM for $\alpha=5,\,\beta=5$, obtained with the non-orthogonal CD expansion, showing the same $1/n$ decay for the first three cumulants.}
    \label{fig:cumulant_nonorth_LRKM}
\end{figure*}
For obtaining the corresponding coefficients, we enforce orthogonality in the momentum sum by inserting the derivative $\partial_k\theta_k$,
\begin{eqnarray}    
&&\sum_k\sin(n\theta_k)q_k(t)\partial_k\theta_k=-\frac{1}{\sqrt L}\sum_k\sum_m\sin(n\theta_k)\sin(m\theta_k)h_m(g)\partial_k\theta_k\\
&&\approx-\sum_m h_m(g)\frac{\sqrt L}{2\pi}\int_0^\pi\mathrm d\theta \sin(n\theta)\sin(m\theta)=-\frac{\sqrt L}{2\pi} h_n(g)\nonumber\\
    &&\Rightarrow h_m(g)=-\frac{\pi}{\sqrt L}\sum_k \frac{\sin(m\theta_k)\sin\theta_k}{1-2g\cos\theta_k+g^2}\partial_k\theta_k\nonumber\\
    &&\approx-\sqrt L\int_0^\pi\mathrm d\theta\frac{\sin(m\theta)\sin\theta}{1+g^2-2g\cos\theta}\approx-\frac{\pi\sqrt L}{4}\begin{cases}
        g^{m-1},\qquad \lvert g\rvert<1,\\
        g^{-(m+1)},\qquad\lvert g\rvert >1,
    \end{cases}\nonumber
\end{eqnarray}
which entirely agrees with the result for the TFIM in the thermodynamic limit~\cite{Damski2014Counterdiabatic}.
Consequently, the small $k$ limit of the phase integral can be expressed as
\begin{eqnarray}
    \int_0^T\mathrm{d}t\, q^{(n)}_k(t)&=&\sum_{m=1}^n \sin(m\theta_k)\int_0^{g_0}\mathrm{d}g\, h_m(g)\\
    &=&\sum_{m=1}^n \sin(mk^z) \left(\int_0^{1}\mathrm dg\,g^{m-1}+\int_1^{g_0}\mathrm dgg^{-m-1}\,\right)=\frac{1}{2}\sum_{m=1}^n\frac{\sin(mk^z)}{m}\nonumber\\
    &=&\frac{\pi-k^z}{2}-\frac{1}{2}\sum_{m=n+1}^\infty\frac{\sin(mk)}{m}\approx\frac{\pi-k^z}{2}-\int_1^\infty\frac{\sin(\kappa x)}{\kappa x}=\frac{\pi-k^z}{2}-\left(\frac{\pi}{2}-\mathrm{Si}(\kappa)\right),\nonumber
\end{eqnarray}
where the Fourier series identity, $\sum_{m=1}^\infty\frac{\sin(mk^z)}{m}=\frac{\pi-k^z}{2}$ was employed and we changed to the rescaled momentum of $\kappa=k^z n$. From here, it is straightforward to compute the excitation probabilities as
\begin{eqnarray}\label{eq: cos_Si}
    p_k&&=\left\lvert\cos\frac{\theta_k}{2}\cos\left[\frac{\pi-k^z}{2}-\left(\frac{\pi}{2}-\mathrm{Si}(\kappa)\right)\right]-\sin\frac{\theta_k}{2}\sin\left[\frac{\pi-k^z}{2}-\left(\frac{\pi}{2}-\mathrm{Si}(\kappa)\right)\right]\right\rvert^2\\
    &&=\left\lvert\cos\frac{\theta_k}{2}\cos\left[\mathrm{Si}(\kappa)-k^z/2\right]-\sin\frac{\theta_k}{2}\sin\left[\mathrm{Si}(\kappa)-k^z/2\right]\right\rvert^2\approx\cos^2\left[\mathrm{Si}(nk^z)\right]+O(1/n).\nonumber
\end{eqnarray}
As a result, the total cumulant generating function can be expressed as
\begin{eqnarray}
    \log\tilde P^{(n)}(\theta;0)&=&\sum_k\log\left[1+(e^{i\theta}-1)p_k\right]\approx\frac{L}{2\pi}\int_0^\pi\mathrm dk\log\left[1+(e^{i\theta}-1)p_k\right]\nonumber\\
    &\propto&n^{-1/z}\int_0^\infty \mathrm dx\log\left[1+(e^{i\theta}-1)\cos^2\left[\mathrm{Si}(x)\right]\right],\qquad
\end{eqnarray}
as this integral is well defined due to the limiting value of $\lim_{x\rightarrow\infty}\mathrm{Si}(x)=\pi/2$, all cumulants will follow the same universal power law of $\kappa_q\propto n^{-1/z},\quad q=1,2,3,\dots$\\
The proposed scaling is verified by the numerical results on the LRKMs in Fig.~\ref{fig:cumulant_nonorth_LRKM} for the short-range regime for $\alpha,\,\beta=5$, studied in detail in the main text, Eq.~\eqref{eq: LRKM_Ham}.

\subsection*{CD expansion in the slow and adiabatic limit}\label{app: cumulants_correction_slow}
In this section, we show details of the analytical approximations for the transition probabilities in the independent TLS of the TFIM in the presence of CD. By this, we demonstrate the CD fast-quench breakdown and the effect of CD in slow quenches.
First, we realize that the results of Ref.~\cite{Damski2014Counterdiabatic},
\begin{eqnarray}
    q^{(n)}_{k}=-4\,\dot g(t)\sum_{m=1}^n\sin(km)\frac{g^{2m}+g^L}{8g^{m+1}(1+g^L)}-\frac{\delta_{m,0}}{8g}, 
\end{eqnarray}
can be expressed in a more compact form.
As the $g^L$ terms will either dominate ($\lvert g\rvert>1$) or disappear ($\lvert g\rvert<1$), $q^{(n)}_k$ can be written as a geometric sum,
\begin{eqnarray}\label{eq: q_CD}
        q^{(n)}_k&\approx&\frac{i\dot g(t)}{4}\sum_{m=0}^{n-1}e^{ik}e^{ikm} g^{\mathrm{sgn}(1-\lvert g\rvert)m}-e^{-ik}e^{-ikm} g^{\mathrm{sgn}(1-\lvert g\rvert)m}\\
    &=&-\dot g\begin{cases}
    &\frac{\sin k-\sin(k(n+1))g^{n}+\sin(kn)g^{n+1}}{2(1+g^2-2g\cos k)},\quad \lvert g\rvert<1,\\
    &\frac{\sin k-\sin(k(n+1))g^{-n}+\sin(kn)g^{-(1+n)}}{2(1+g^2-2g\cos k)},\quad \lvert g\rvert>1.
    \end{cases}\nonumber
\end{eqnarray}

 To clarify how the dynamics is modified, we focus on the avoided crossings that would arise in the absence of CD, around the approximate position, $g\approx1-k^2/2$. For modes $k\ll1/n$ and for time scales $T\gg n^2$, excitations happen with finite Landau-Zener (LZ) probabilities whenever $Tk^2\sim O(1)$. In this regime, however, the CD term scales as $q^{(n)}\sim nk/(k^2T)\ll1$ and thus the dynamics becomes insensitive to it. For larger momenta, adiabaticity is instead enforced by CD. This can be understood from the fact that already in sudden quenches, the diabatic effects of the bare Hamiltonian appear only at subleading order, and adiabaticity is ensured to leading order by the CD. Slower quenches merely suppress further these subleading corrections, leaving the leading-order behavior unchanged.

 \begin{figure}[b!]
\centering

    \includegraphics[width=.6\linewidth]{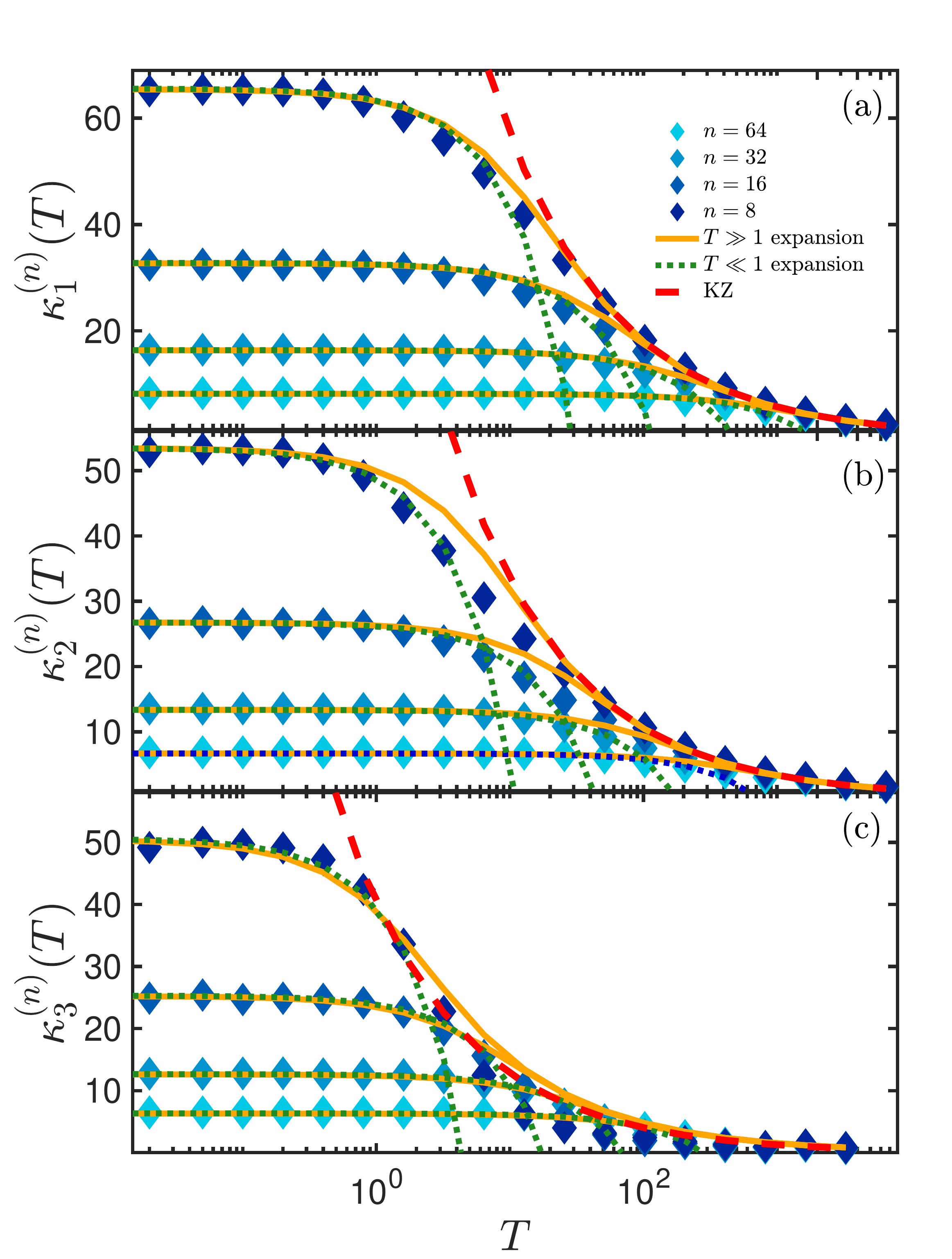}
    \caption{$(a)$ Mean, $(b)$ variance, and $(c)$ skewness of the defect statistics for CD expansion orders $n=8,16,32,64$ and $L=1600$. The first-order correction around the sudden limit provides a slightly more accurate description of the intermediate CD regime than the analytical ansatz derived from the slow-driving limit.}
    \label{fig:cumulants}
\end{figure}
 
 For $T\ll n^2$, the low modes exhibit a fast-quench plateau from the LZ perspective, converging to the limiting value of the CD-assisted sudden-quench. For larger momentum $k\gtrsim1/n$, the numerator of the CD remains finite, while its denominator vanishes as $k^2 T\ll 1$ according to the suppression of the higher momentum LZ transitions in the fast-quench plateau, making the CD term dominant over the bare process and so ensuring adiabatic dynamics again.
Altogether, this approximate picture indicates that, to leading order, local CD enforces adiabatic evolution for high-momentum modes, whereas low-momentum modes ($k \lesssim k_n \propto 1/n$) experience subleading LZ excitations unaffected by the control.
As a result, the cumulant generating function admits  an analogous form to the one in the bare process but with the LZ probabilities restricted below the momentum $k_n\propto1/n$,
\begin{eqnarray}\label{eq: logP_CD_Krylov}
    \log\tilde P^{(n)}(\theta;T)=\sum_{k<k_n}\log\left[1+(e^{i\theta}-1)p_k\right]
    =-L\,n_\mathrm{ex}\sum_{p=1}^\infty\,\frac{(1-e^{i\theta})^p}{p^{3/2}}\mathrm{erf}\left(\frac{k_n\sqrt{p}}{2\sqrt\pi n_\mathrm{ex}}\right),\nonumber
\end{eqnarray}
where $n_\mathrm{ex}=(8\pi^2T)^{-1/2}$ is the average defect density without CD.
For $k^2_nT\gg1$, i.e. $T\gg T^\mathrm{CD}_\mathrm{fast}$ the correction becomes exponentially close to the result of Ref.~\cite{delCampo18}, $\mathrm{erf}\left[\frac{k_n\sqrt{p}}{2\sqrt\pi n_\mathrm{ex}}\right]\approx 1$. However, decreasing the driving time beyond the fast-quench breakdown, the average extracted from Eq.~\eqref{eq: logP_CD_Krylov} converges to a constant value,
\begin{eqnarray}
    \kappa^{(n)}_1(T)=-i\partial_\theta\log\tilde P^{(n)}(\theta;T)\Big\vert_{\theta=0}=Ln_\mathrm{ex}\,\mathrm{erf}\left(\frac{k_n\sqrt{p}}{2\sqrt\pi n_\mathrm{ex}}\right)\rightarrow \frac{k_n}{\pi}\,L.
\end{eqnarray}
This allows for fixing the constant factor in $k_n$ by requiring that $n_\mathrm{ex}\mathrm{erf}\left[\frac{k_n}{2\sqrt\pi n_\mathrm{ex}}\right]\approx 1.05/\pi\,n^{-1}$ leading to $k_n\propto 1.05\,n^{-1}$ so that it reproduces the sudden-quench limit.

Using this argument as an ansatz, the analytical structure of all cumulants can be captured approximately around the CD fast-quench breakdown, providing a smooth function interpolating between the exact results in the fast and slow driving regimes.
Similarly to the average, we restrict ourselves to the first-order expansion of Eq.~\eqref{eq: logP_CD_Krylov} and fix the constant $k_n$ so that the cumulants reach the exact CD fast-quench values for $T\rightarrow 0$. As a result of the limiting form of the error function, $\mathrm{erf}(x)=\frac{2}{\sqrt\pi}x+O(x^3)$, the approximation correction from the slow driving limit is given by
\begin{equation}\label{eq:kappa_n_T}
    \kappa^{(n)}_q(T)\approx L\,\kappa^\mathrm{KZ}_q(T)\mathrm{erf}\left[\frac{\sqrt\pi\kappa^{(n)}_q(0)}{2\kappa^\mathrm{KZ}_q(T)}\right]=L\,\kappa^\mathrm{KZ}_q(T)\frac{2}{\sqrt\pi}\frac{\sqrt\pi\kappa^\mathrm{(n)}_q(0)}{2\kappa^\mathrm{KZ}_q(T)}+O(T^{3/2})\approx \kappa^{(n)}_q(0),
\end{equation}
where $\kappa^\mathrm{KZ}_q(T)$ and $\kappa^{(n)}_q(0)$ denote the $q$-th cumulant for the $n$-th order of expansion in the KZ scaling and sudden limits, respectively. Here, the last approximation shows that the leading order with respect to $T\rightarrow 0$ properly recovers the sudden-quench limit.

Note that this strategy replaces the approach of the $1/T$ expansion reported in Ref.~\cite{Grabarits2025}.  In particular, without CD, the LZ formula only applies up to momentum $k\lesssim T^{-1/4}$. According to the results of Ref.~\cite{Grabarits2025}, beyond the CD fast-quench regime, $n^2\ll T$, this implies a scale of $k\lesssim n^{-1/2}$. Thus, the KZ scaling regime beyond the CD fast-quench threshold is completely governed by the LZ transitions.
The numerical verification of these predictions is shown in Fig.~\ref{fig:cumulants}, Eq.~\eqref{eq:kappa_n_T} provides an excellent fit to the numerical results in the crossover regime between the fast and slow limits.

The universal fast-quench cumulant scaling, $\kappa^{(n)}_1(0)\approx\frac{1.05}{\pi}n^{-1}$ with the prefactor determined numerically in the TFIM, also reveals additional features of the adiabatic regime. The onset of adiabaticity can be captured by $\kappa^{(n)}_1(0)= 2/L$, defining the adiabatic threshold value of the CD expansion order, $ n^\mathrm{CD}_\mathrm{ad}\approx\frac{1.05}{2\pi}\,L$.
Thus, adiabaticity emerges when the dynamics is assisted by a local CD expansion of significantly lower order than that of the exact CD protocol, $n=L$. This justifies the efficiency of approximate local CD protocols.
Above this threshold, defect formation is restricted to the zero mode in the leading order, and the ground-state fidelity, otherwise exponentially small with $L$, becomes finite. As a result, the CD fast-quench plateau is then equivalent to a quasi-adiabatic regime, while the CD fast-quench breakdown results in reaching the adiabatic limit $T^\mathrm{CD}_\mathrm{ad}\approx T^\mathrm{CD}_\mathrm{fast}$.

 \smallskip

\noindent{\textbf{\textsf{Data availability}}}\\
The data supporting the findings of this study can be found via Zenodo, 

\smallskip

\noindent{\textbf{\textsf{Code availability}}}\\
The codes evaluating the generated data in this study can be found via Zenodo, 

\smallskip

\noindent{\textbf{\textsf{Acknowledgements}}}\\
We are grateful to Federico Balducci for insightful discussions.
This project was supported by the Luxembourg National Research Fund  C22/MS/17132054/AQCQNET, and CC25/MS/19559370/FastQOPT.

\smallskip

\noindent{\textbf{\textsf{Author contributions}}}\\
 A. G.  and A. d. C. provided the corresponding analytical background. All authors contributed to the writing of the manuscript.

\smallskip
\noindent{\textbf{\textsf{Competing interests}}}\\
 The authors declare no competing financial or non-financial interests.

\bibliography{references}

\end{document}